\documentclass[namedreferences]{solarphysics}

\usepackage[hyperref,optionalrh]{spr-sola-addons} 
\usepackage{graphicx}        
\usepackage{color}           
\usepackage{breakurl}        
\usepackage{url}
\usepackage{natbib}

\newcommand{\so}{    {SOL2012-03-13}} 
\newcommand{\event}{SOL2012-03-13}
\newcommand{\FLA}{{FLA}}
  
\newcommand{\FO}{{F1}}
\newcommand{\FT}{{F2}}
\newcommand{\CS}{{C7}}

\newcommand\arcsec{\mbox{$^{\prime\prime}$}}

\begin{document}

\begin{article}

\begin{opening}

\title{Origin of the 30 THz emission detected during the 2012 March 13 solar flare at  17:20 UT\\ {\it Solar Physics}}

\author{G.~\surname{Trottet}$^{1}$\sep
        J.-P.~\surname{Raulin}$^{2}$\sep
       A.~\surname{Mackinnon}$^{3}$\sep     
 G. ~\surname{Gim\'enez de Castro}$^{2}$\sep
P.J.A. ~\surname{Sim\~oes}$^{3}$\sep
D. ~\surname{Cabezas}$^{2}$\sep
V. ~\surname{de La Luz}$^{4}$\sep
M. ~\surname{Luoni}$^{5}$\sep
P. ~\surname{Kaufmann}$^{2,6}$\sep
      }
\runningauthor{Trottet et al.}
\runningtitle{Origin of the 30 THz emission during a solar flare}

   \institute{$^{1}$ LESIA, Observatoire de Paris, PSL Research University, CNRS, Sorbonne UniversitŽs, UPMC Univ. Paris 06, Univ. Paris Diderot, Sorbonne Paris 			CitŽ, 5 place Jules Janssen, 92195 Meudon, France;
  		email:~\url{gerard.trottet@obspm.fr}\\ 
		 $^{2}$ CRAAM Universidade Presbiteriana Mackenzie, S\~ao Paulo, Brazil;
		email:~\url{raulin@craam.mackenzie.br};
		email:~\url{guigue@craam.mackenzie.br};
		email:~\url{kaufmann@craam.mackenzie.br};
		email:~\url{deniscabezas@gmail.com}\\
		$^3$ School of Physics and Astronomy, SUPA, University of Glasgow, Glasgow G12 8QQ, UK;
		email:~\url{Alexander.MacKinnon@glasgow.ac.uk};
		email:~\url{paulo.simoes@glasgow.ac.uk}\\
		$^{4}$ SCiESMEX, Instituto de Geofisica, Unidad Michoacan, Universidad
		Nacional Autonoma de Mexico, Morelia, Michoacan, Mexico. CP 58190;
		email:~\url{itztli@gmail.com}\\
		$^{5}$ IAFE, University of Buenos Aires, Buenos Aires, Argentina;
		email:~\url{mariluoni@gmail.com}\\
		$^{6}$ CCS, University of Campinas, Campinas, Brazil
            }

\begin{abstract}
Solar observations in the infrared domain can bring important clues on the response of the low solar atmosphere to primary energy released during flares. At present the infrared continuum has been detected at 30 THz (10 $\mu$m) in only a few flares. SOL2012-03-13 , which is one of these flares, has been presented and discussed in Kaufmann et al. (2013). No firm conclusions were drawn on the origin of the mid-infrared radiation. In this work we present a detailed multi-frequency analysis of the SOL2012-03-13 event, including observations at radio millimeter and sub--millimeter wavelengths, in hard X-rays (HXR), gamma-rays (GR),  H$\alpha$, and white-light. HXR/GR spectral analysis shows that \so\ is a GR line flare and allows estimating the numbers of and energy contents in electrons, protons and $\alpha$ particles produced during the flare. The energy spectrum of the electrons producing the HXR/GR continuum  is consistent with a broken power-law with an energy break at $\sim$ 800 keV. It is shown that the high-energy part (above $\sim$ 800 keV) of this distribution is responsible for the high-frequency radio emission ($>$ 20 GHz) detected during the flare. By comparing the 30 THz emission expected from semi-empirical and time-independent models of the quiet and flare atmospheres, we find that most ($\sim$80\%) of the observed 30 THz radiation can be attributed to  thermal free--free emission of an optically-thin source. Using the F2 flare atmospheric model \citep{Mac:al-80} this thin source is found to be at temperatures T $\sim$ 8000 K and is located well above the minimum temperature region.  We argue that the chromospheric heating, which results in 80 \% of the 30 THz excess radiation, can be due to energy deposition by non-thermal flare accelerated electrons, protons and $\alpha$ particles. The remaining 20\% of the 30 THz excess emission is found to be radiated from an optically-thick atmospheric layer at T $\sim$ 5000 K, below the temperature minimum region, where direct heating by non-thermal particles is insufficient to account for the observed infrared radiation.
\end{abstract}
\
\end{opening}

\section{Introduction}
     \label{S-Intro} 
Solar flare continuum observations from the mid-infrared domain (a few tens of THz) to the millimeter--sub-millimeter radio domain provide in principle unique diagnostics of energy transport processes from the flare energy release region to the chromosphere and of the most energetic flare accelerated particles \citep[\textit{e.g.}][and references therein]{Tro:Kle-13}. Since year 2000, radio observations at 212 and 405 GHz have been routinely obtained by the \textit{Solar Submillimeter Telescope} (SST) and more recently high-cadence imaging observations of both the quiet Sun and flares have been obtained at 30 THz \citep[][ and references therein]{Kau:al-08}. The 2012  March 13 flare at $\sim$ 17:22 UT (\so) was the first flare observed at 30 THz \citep{Kau:al-13}. Since then other flares detected at 30 THz have been reported in the literature \citep[][and references therein]{Kau:al-15}. During \so\, the time evolution of the 30 THz impulsive  emission was found to be similar to that of the radio (1-212 GHz), hard X-ray and white-light emission. The 30 THz emitting source was also co-spatial with a region of flare-enhanced white light continuum. No firm conclusion on the origin of the 30 THz infrared emission observed during \so\ was drawn by  \cite{Kau:al-13}. Nevertheless, they suggested  that it is consistent with heating of the flaring atmosphere below the temperature minimum region which cannot be due to direct heating by hard X-ray emitting electrons, at least within the classic thick-target model. On the other hand, theoretical work by \cite{Okh:Hud-75} raised the possibility that the infrared continuum at wavelengths shorter than $\sim$ 20 $\mu$m may arise from an optically-thin source at temperatures in the range of 10$^4$--10$^5$ K. This latter possibility is in line with the results of time-dependent simulations of the chromospheric heating by electron beams performed by \cite{Kas:al-09b}. 

The main goal of this paper is to investigate what is the origin of the infrared continuum  detected at 30 THz during \so. For that we have performed a detailed analysis of multi-frequency observations including measurements in radio millimeter and sub millimeter wavelengths, hard X-ray (HXR) and gamma-ray (GR),  H$\alpha$, and white-light (WL). The instrumentation used to collect these data is briefly described in Section~\ref{S-Inst} and an overview of the observations is presented in Section~\ref{S-Over}. Section~\ref{S-xrgr} is devoted to the analysis of HXR/GR spectra from which  numbers and energy contents of electrons, protons and $\alpha$ particles accelerated during \so\ are estimated. In Section~\ref{S-rad} we discuss the relationship between HXR/GR and radio emitting electrons.  The origin of the 30 THz emission is examined in Section~\ref{S-30T} by using semi-empirical time-independent models of both the quiet and flare chromosphere. In Section~\ref{S-Resp} we provide an estimate of the energy deposited by flare accelerated electrons, protons and $\alpha$ particles in the atmospheric layers which radiate the 30 THz emission and we examine if it can account for the energy radiated at 30 THz. The final Section summarizes the main conclusions.
\section{Instrumentation} 
      \label{S-Inst}      

\begin{figure}    
 \centerline{\includegraphics[width=\textwidth,clip=]{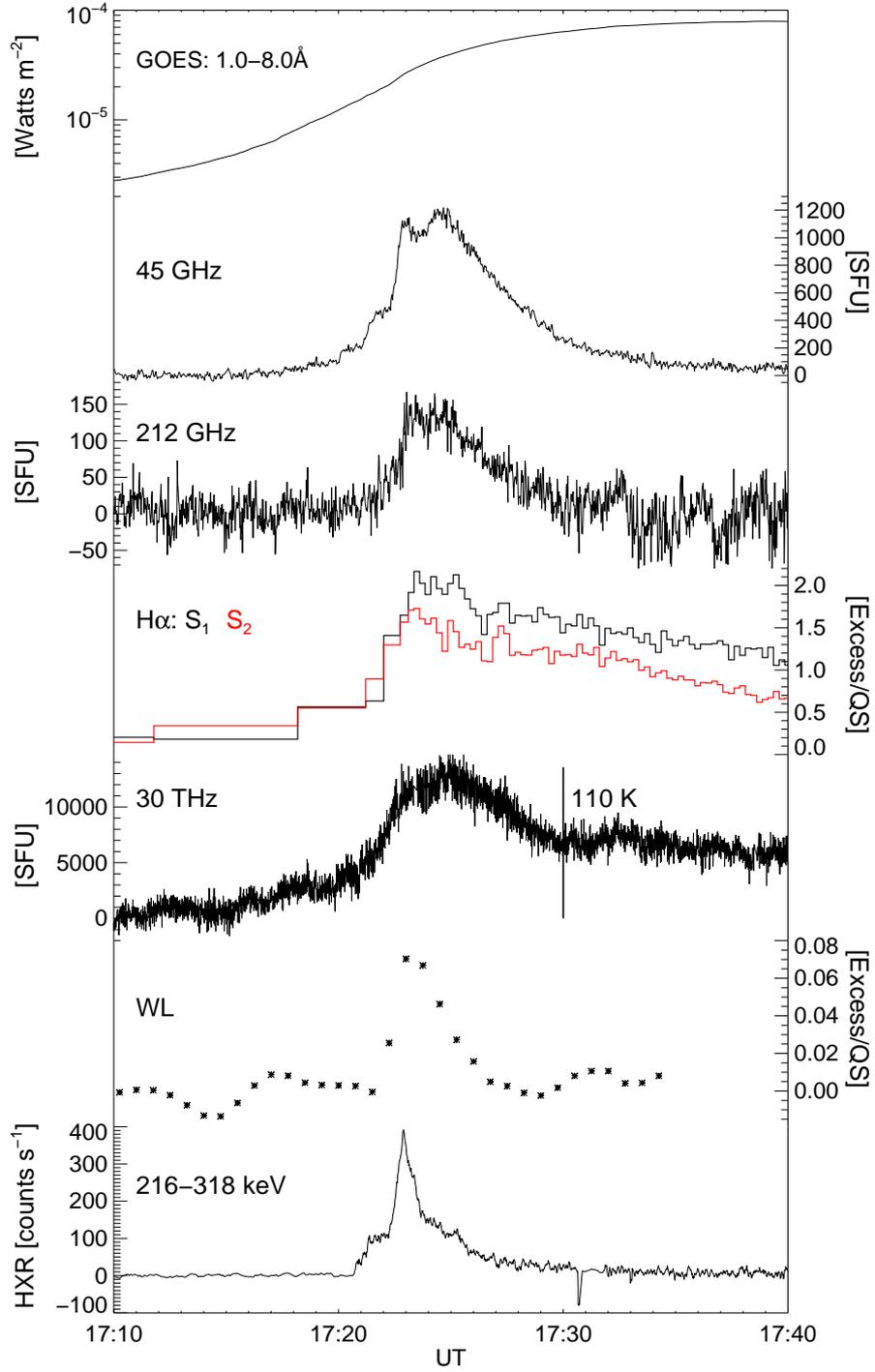}
              }
              \caption{ Temporal evolution  of the flare SOL2012-03-13  in various wavelength domains }
   \label{Fig_over} 
   \end{figure}
Radio data at 45 GHz and  90 GHz  were obtained by the radio polarimeters described in \cite{Val:al-13}. For the \so\ flare the uncertainty on flux densities is about 20\%.  Observations at  212  and 405 GHz have been provided by the SST.  The atmospheric opacity was estimated to be 0.47 and 2.2 Nepers, at 212 and 405 GHz respectively. At 212 GHz the flux density accuracy is estimated to be 20\%. At 405 GHz the 1 $\sigma$ detection threshold was $\sim$ 55 sfu\footnote{$\rm{1sfu = 10^{-22}W\cdot m^{-2}\cdot Hz^{-1}}$}, and no significant emission was detected.

In the HXR and GR domain, the  \so\ flare has been detected by the \textit{Gamma-Ray Burst Monitor} (GBM) on \textit{Fermi} \citep{Mee:al-09}.  GBM is composed of twelve sodium-iodide (Na I) and two bismuth germanium oxide (BGO) detectors. In the present analysis we have used the most sunward Na I and BGO detectors. The Na I data consist in 128 energy channels covering the energy range 4 keV--2000 keV and  the BGO data also consist of 128 energy channels but in the 113--50,000 keV range.

The 30 THz instrumental setup at El Leoncito is reported in \cite{Mar:al-08} and 
\cite{Kau:al-08}. For the present solar flare the data calibration and the determination of the excess flux 
density relative to quiet Sun have been described in details in \cite{Kau:al-13}. The accuracy of flux density measurements has been estimated to be $\sim$ 25\%.  The diffraction limit of the instrument is $\sim$ 15\arcsec\ and the source position uncertainty is $\sigma \sim$ 3\arcsec\ as obtained from fitting of the 30 THz limb. The bandwidth of the instrument is $\sim$ 17 THz and in the following, we assume that the 30 THz emission does not depend on frequency within this bandwidth.

H$\alpha$ observations were obtained with the \textit{H-Alpha Solar Telescope 
for Argentina} \citep[HASTA; ] []{Bag:al-99}.
HASTA provides solar full disk images with a time resolution of 5 seconds in flare mode and a spatial resolution of 2\arcsec.
Excess emission due to the flare has been normalized to the quiet Sun emission.

The above data set has been complemented by radio observations at 1.415, 2.695, 4.995 and 8.8 GHz from the 
Sagamore Hill-station of the  US Air Force {\it Radio Solar Telescope Network} (RSTN)\footnote{\url{http://www.ngdc.noaa.gov/stp/space-weather/solar-data/solar-features/solar-radio/rstn-1-second/}}  \citep{Gui-79}. During \so\ there were no reliable 15.4 GHz measurements. We have also used WL images from the \textit{Helioseismic and Magnetic Imager} \citep[HMI;][]{She:al-12} on board the {\it Solar Dynamic Observatory} (SDO) obtained every 45 s with a spatial resolution better than  $\sim$ 1 \arcsec\  \footnote{\url{http://hmi.stanford.edu/Description/HMI_Overview.pdf}}.

\section{Observation overview} 
      \label{S-Over}  

 \begin{figure}    
 \centerline{\includegraphics[width=\textwidth]{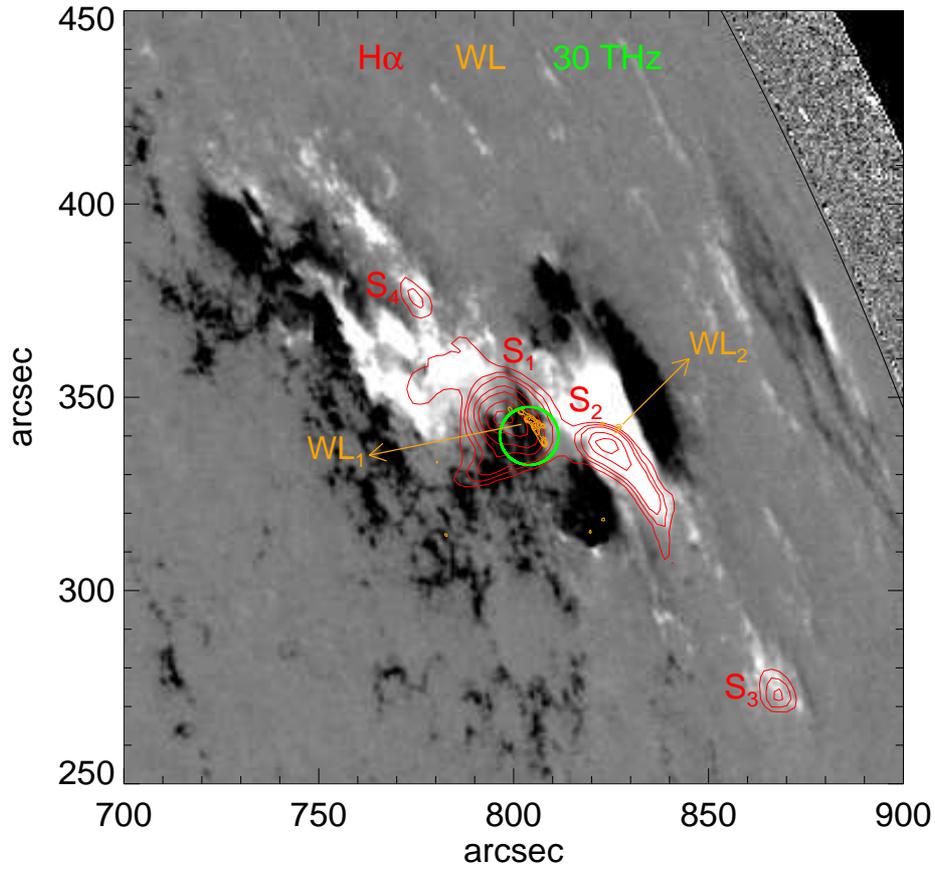}
              }
 \caption{Projections of H$\alpha$ (17:22:53  UT, red contours), WL (17:23:00  UT, yellow contours) and 30 THz (17:22:54  UT, green circle) sources on a SDO/HMI vector magnetogram obtained at 17:13:15 UT}
   \label{Fig_images} 
   \end{figure}

The flare \so\ occurred in NOAA active region AR 11429 located at N18W64 on 2012 March 13 at 22:00 UT. 
It was associated with a M7.9 soft X-ray (SXR) event detected by the {\it Geostationary Operational Environmental Satellite} ({\it GOES}) and with a H$\alpha$ flare of importance 1B. From top to bottom Fig.~\ref{Fig_over} displays the time profile of the SXR (1-8 \AA) from {\it GOES}\footnote{provided by NASA/GSFC at \url{http://umbra.nascom.nasa.gov/goes/fits/}},  radio (45 and 212 GHz), H$\alpha$, flare excess mid-infrared (30 THz), WL, and  216-318 keV HXR emissions detected during \so. Except  H$\alpha$ measurements, these observations have been presented  in \cite{Kau:al-13}. The HXR light curve consists in an impulsive burst which started at $\sim$ 17:20:53 UT with a maximum at $\sim$ 17:22:54 UT and lasted for about ten minutes.  This impulsive peak is seen from radio to optical wavelengths,  i.e. from the corona down to the low chromosphere. At radio wavelengths, there are two peaks of apparently similar flux densities. There is a change of the decay slope of the  HXR emission which corresponds to the second radio peak. The H$\alpha$, 30 THz and WL emissions show a broad peak which encompasses the two radio and HXR peaks. In addition to this impulsive phase, there is a long-lasting and slowly varying component seen at all wavelengths except in HXR and WL. For example at 45 GHz, although it is weak ($\sim$ 40-50 sfu), the gradual emission is still visible after 19:00 UT.

Figure~\ref{Fig_images} shows the projections of the H$\alpha$, WL and 30 THz sources, obtained around the maximum of the impulsive peak, on a vector magnetogram recorded by SDO/HMI at 17:13:15 UT. The flare enhanced H$\alpha$ emission arises from four sources (red contours) marked S$_1$, S$_2$, S$_3$ and S$_4$.  S$_1$ and S$_2$ are close to the two WL emitting regions, WL$_1$ and WL$_2$, shown by yellow contours \citep[see also Figures ~3 and 10 in ][]{Kau:al-13}.  The larger extents of  WL$_1$ and WL$_2$ are respectively $\sim$ 10\arcsec\ and 2\arcsec. The projections of  WL$_1$ and WL$_2$ and S$_1$ and S$_2$ are located on opposite magnetic field polarities and are connected by a bright loop shown at 211 \AA\ in Figure 5 of \cite{Kau:al-13}. At 30 THz  only the source  corresponding to S1 and to WL$_1$, the stronger WL emitting region,  is detected within the sensitivity of the present setup of the 30 THz instrument. Since the 30 THz source is not resolved by the instrument, it is marked by a green circle whose diameter is the diffraction limit of the instrument (15\arcsec) in both directions. Figure~\ref{Fig_images} thus indicates that the 30 THz source and the WL barycenter emission, which is close to maximum of WL$_1$,  are located close to each other. However, the uncertainty on the 30 THz source position (3 $\sigma \sim$ 9\arcsec) 
prevents to draw any conclusion on  the relative height of both sources, for relative heights lower than  few thousands of km. While the time profiles of S$_1$ and S$_2$  show a clear impulsive phase counterpart (see Figure~\ref{Fig_over}), the remote sources S$_3$ and S$_4$ only show a slowly varying emission that will not be considered  in the following.

\section{Electron and ion numbers and energy contents around the maximum of \so}
     \label{S-xrgr}   

  \begin{figure}    
 \centerline{\includegraphics[width=\textwidth,clip=]{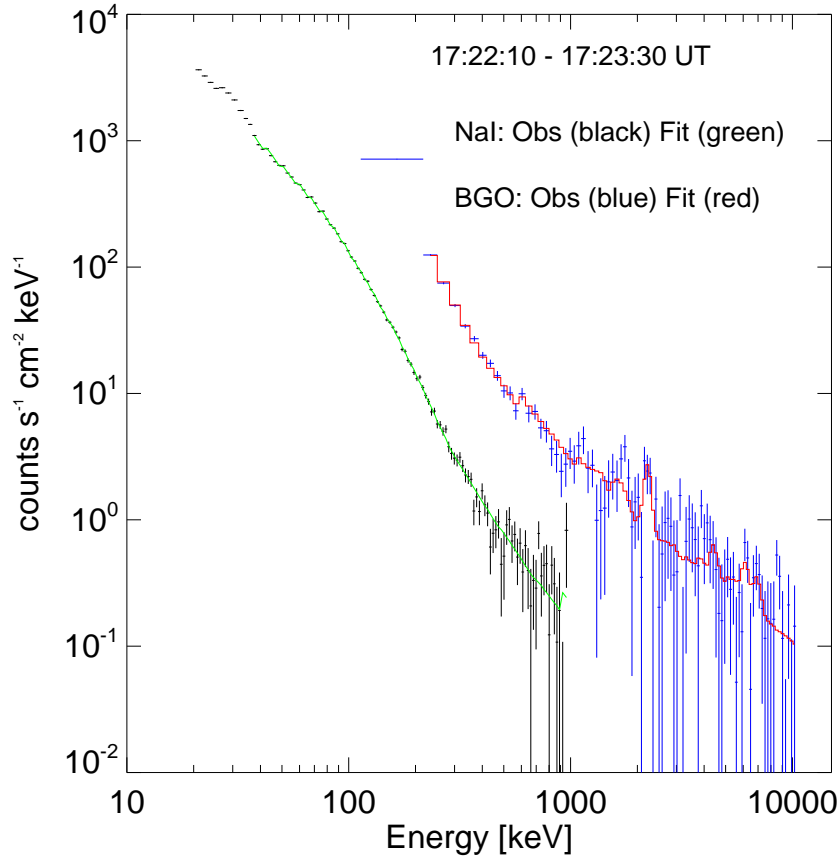}
              }
              \caption{ HXR/GR spectrum  around the maximum of SOL2012-03-13  obtained by combining 
              BGO and Na I measurements of the GBM instrument on \textit{Fermi} (error bars). The  green and red curves 
              show the model spectrun fitted to GBM observations (see text).}
   \label{Fig_ferm}
   \end{figure}

Figure~\ref{Fig_ferm} shows the mean excess count rate spectrum in the 20-10,000 keV range recorded by the most sunward NaI (black error bars) and the most sunward BGO (blue error bars) detectors, for a 80 s time interval around the peak of the HXR/GR emission. In the 36-1000 keV range, the observed NaI count rate spectrum is well represented by the expected count rate spectrum shown by the green line. This latter spectrum has been obtained  by convolving a triple power-law photon spectrum with the NaI detector calibration matrix. The triple power-law spectrum represents the electron bremsstrahlung spectrum in the $\sim$ 35-1000 keV range. It is defined by the normalization factor at 50 keV, the power-law indices $\gamma_1$, $\gamma_2$ and  $\gamma_3$ and the break energies Ebr$_1$ and Ebr$_2$ displayed in Table~\ref{T1}.  
In  the 216-10,000 keV range, the BGO count rate spectrum is well represented by the expected count rate spectrum (red line) obtained for a photon spectrum which is the sum of a broken power-law,  of the 2.2 MeV neutron capture line and of a template of narrow nuclear lines.   We used the nuclear line template, included as standard within OSPEX \citep{Sch:al-02}, calculated for a flare  at  an heliocentric angle of $60^\circ$  by assuming a downward isotropic distribution of ions with a power-law  energy distribution  of index 4 and an $\alpha$/p ratio of 0.22. 
The broken power-law represents the electron bremsstrahlung spectrum in the 200-10,000 keV range. It is defined by the normalization factor at 50 keV, the power-law indices $\Gamma_1$ and $\Gamma_2$ and the break energy Ebra (see Table~\ref{T2}). For the fitting procedure we have used the OSPEX package of Solar Soft.

 \begin{table}   
\caption{ Parameters of the photon spectrum fitted to the NaI measurements in the 35-1000 keV
}
\label{T1}
\begin{tabular}{cccccc}     
  \hline                   
  photons/s/cm$^2$/keV & $\gamma_1$ & Ebr$_1$ & $\gamma_2$ &  Ebr$_2$ & $\gamma_3$  \\
  at 50 keV &   &  keV &  & keV &  \\
  \hline
4.7 $\pm$ 1.5 & 3.3$\pm$ 0.2 & 94 $\pm$ 10 & 3.8 $\pm 0.2$ & 282 $\pm$ 20 & 2.4 $\pm$ 0.2\\
\end{tabular}
\end{table}
  
\begin{table}   
\caption{ Parameters of the photon spectrum fitted to the BGO measurements in the 200-10,000 keV
}
\label{T2}
\begin{tabular}{cccccc}     
  \hline                   
photons/s/cm$^2$/keV   & $\Gamma_1$ & Ebra & $\Gamma_2$ &  2.2 MeV line & narrow lines  \\
  at 50 keV &   &  keV &  & photons/s & photons/s  \\
\hline
5.7 $\pm$ 1.9 & 3.6 $\pm$ 0.2 & 388 $\pm$ 50 & 2.3 $\pm$ 0.2 & 0.054 $\pm$ 0.018 & 0.11 $\pm$ 0.04\\
 \hline
  electrons/s/keV & $\delta_1$ & Ebre & $\delta_2$ & & \\
  at 50 keV & & [keV] & & &\\
  \hline
 $ 5.6\times10^{33}$ & 4.7 & 800 & 3.5 & & \\
\end{tabular}
\end{table}

 Figure~Ê\ref{Fig_ferm} shows that: (i) the electron bremsstrahlung continuum extends up to at least 10,000 keV and (ii) there is clear 2.2 MeV line emission and weak narrow GR line emission indicating that $\sim$ 1-100 MeV protons and $\sim$ 1-100 MeV/nucleon ions have been accelerated during \so.  The photon fluxes derived from both Na I and BGO data agree well (within better than 25\%) between 200-700 keV. However, because the responses of NaI and BGO detectors are quite different, a given incident photon spectrum will produce different count spectra in both types of detectors as shown in Figure~Ê\ref{Fig_ferm}.  Above $\sim$ 400 keV, the 1-$\sigma$ uncertainty is smaller for BGO than for Na I.  In the following we will only consider the photon spectrum derived from BGO data since we are interested in microwave emitting electrons and in GR line radiation. In order to derive the spectrum of electrons from this photon spectrum we further assume that this electron spectrum is also a broken power law. The electron spectrum is then derived  by using an electron-to-thick-target bremsstrahlung code including both relativistic effects and electron-electron bremsstrahlung\footnote{N. Vilmer, private communication} and by varying its parameters  until a good agreement with the photon spectrum is reached. The obtained parameters are displayed in Table~\ref{T2}. The mean flux of $ >$ 50 keV electrons between 17:22:10 and 17:23:30 UT is found to be F$_{\rm {e}} (>50~ \rm{keV}) \sim 8\times 10^{34}$ electrons s$^{-1}$ corresponding to a mean energy flux $\Phi_{\rm{e}}(> 50~ \rm{keV}) \sim 9\times 10^{27}$ ergs s$^{-1}$ . The nuclear line template is normalized so that an amplitude of 1 corresponds to 8.5946 $\times$ 10$^{29}$ protons with energies above 30 MeV.  Here the amplitude of the nuclear line template is 0.11 $\pm$ 0.04 s$^{-1}$ (see Table~\ref{T2}). It is well documented that ions with energies down to almost 1 MeV/nucleon contribute to the GR line emission.  For  the assumed energy power-law index of 4 we then get mean proton and $\alpha$ fluxes at the top of the atmosphere: F$_{\rm{p}}(>\rm{1~MeV}) \sim 2.6\times10^{33}$ protons s$^{-1}$, F$_{\rm{\alpha}}(\rm{>1~MeV/nucl})\sim 5.6\times10^{32}$ $\alpha$ s$^{-1}$. These correspond to energy fluxes of respectively $\Phi_{\rm{p}}(\rm{>1~MeV})=6.2\times10^{27}$ ergs s$^{-1}$ and $\Phi_{\rm{\alpha}}(\rm{>1~MeV}) \sim 5.4\times10^{27}$ ergs s$^{-1}$.
Since the uncertainty on the determination of photon fluxes is about 30\%, the above values of particle and energy fluxes suffer from the same uncertainty. The broad GR line emission has not been taken into account to get the above estimates because, due to the poor statistics, the observed GR line spectrum is not well defined. Nevertheless, we have checked that the inclusion of broad lines lead to estimates of proton and $\alpha$ particle fluxes and energy fluxes that stay within the 30\% uncertainty. It should be noted that $\Phi_{\rm{p}} \sim \Phi_{\rm{\alpha}}$ and that $\Phi_{\rm{e}}(> 50~ \rm{keV}) \sim \Phi_{\rm{p}}(\rm{>1~MeV})+\Phi_{\rm{\alpha}}(\rm{>1~MeV})$.

\section{Relationship between radio and HXR/GR emitting electrons during the impulsive phase} 
\label{S-rad}

     \begin{figure}    
 \centerline{\includegraphics[width=\textwidth,clip=]{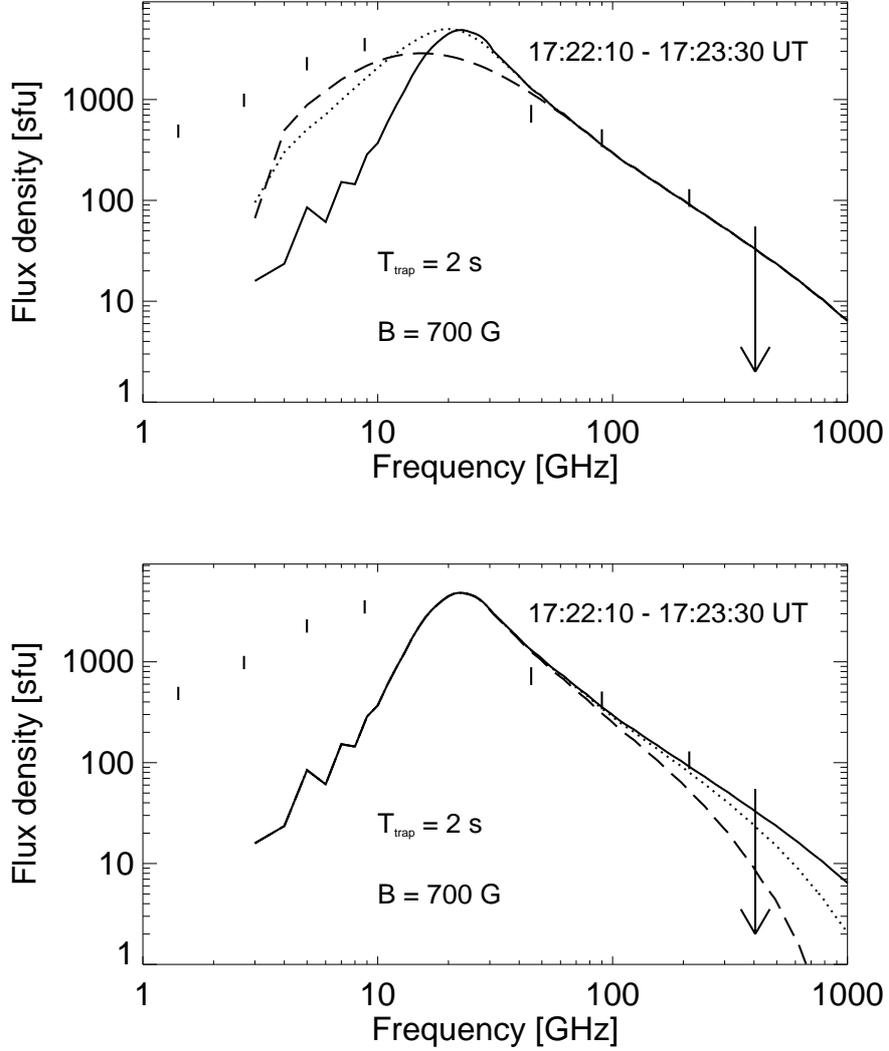}
              }
              \caption{Radio spectrum: in both panels the vertical error bars show the radio spectrum observed in the 1-212 GHz frequency range around the maximum of the impulsive phase of \so\.. At 405 GHz the arrow indicates the upper limit of the flux density. The gyrosynchrotron spectra computed for  $\rm{B=700~G}$ and $\rm{T_{trap}=2~s}$ ,   $\rm{E}_{\rm{max}}=10,000~\rm{keV}$  and $\rm{E}_{\rm{min}}=$  30 keV (solid line), 400 keV (dotted line) and 800 keV (dashed line) are over plotted in the upper panel while results for $\rm{E}_{\rm{min}}=30~Ê\rm{keV}$ and  $\rm{E}_{\rm{max}}=$ 10,000 keV (solid line), 7,000 keV (dotted line) and 4,000 keV (dashed line) are over plotted in the lower panel (see text).}
   \label{Fig_gyro}
   \end{figure}

Figure~\ref{Fig_gyro} displays the averaged radio spectrum (vertical error bars) observed during the same time interval (17:22:10--17:23:30 UT)  as the HXR/GR spectrum analyzed in Section~\ref{S-xrgr}. Its shape is reminiscent of a gyrosynchrotron spectrum with a turnover frequency somewhere between 10 and 30 GHz. It is generally admitted that the 1-200 GHz (cm-mm) radio emission and the HXR/GR continuum are both radiated by a closely related population of electrons accelerated in the corona \citep[\textit{e.g.}][]{Bas:al-98, Pic:Vil-08}. Briefly stated: (i) the gyrosynchrotron emission is radiated by the instantaneous population of electrons present in the coronal portion of magnetic loops connected to the acceleration region, while (ii) the HXR/GR bremsstrahlung continuum is produced  by thick-target interactions of precipitating electrons at the chromospheric foot points of these loops. It is also well documented that for magnetic field of a few hundred Gauss the cm--mm emission is radiated by electrons with energies ranging from a few hundreds of keV to a few MeV \citep[\textit{e.g.}][]{Ram-69, Pick:al-90, Ram:al-94}.  As a first approximation we thus assume that the instantaneous spectrum of radio emitting electrons is given by \citep[\textit{e.g.}][]{Tro:al-98}: $ \rm{N}_{\rm{R}}(\rm{E})\approx F_{\rm{X}}(\rm{E})~\rm{T}_{\rm{trap}}$ electrons keV$^{-1}$ where $\rm{T}_{\rm{trap}}$ is the time spent by the accelerated electrons in the cm--mm radio emitting region and F$_{\rm{X}}(\rm{E)}$ is the electron flux spectrum displayed in Table~\ref{T2}. Without multi-frequency cm-mm imaging observations it is not possible to constrain radio source models with inhomogeneous magnetic field B and ambient density N$_{\rm{amb}}$. In the following we thus compute the gyrosynchrotron emission produced by $ \rm{N}_{\rm{R}}(\rm{E})$  in a radio emitting region of constant  N$_{\rm{amb}}$ and uniform B. Such a simple model is not able to account for the $<$ 10 GHz optically-thick radio emission, but it allows one, in principle,  to reasonably match the optically-thin part of the radio spectrum observed at 45, 90 and 212 GHz during the present flare.  The gyrosynchrotron emission has then been computed for $ \rm{B} = 700 $ G, $\rm{T}_{\rm{trap}}=2~s$ and N$_{\rm{amb}}= 5\times10^{10}~ \rm{cm}^{-3}$. The diameter and thickness of the radio source normal to and along the line of sight have been set to  14,000 and 2000 km respectively and the angle between B and the line of sight to $45^\circ$.  The gyrosynchrotron code used is that by \cite{Ram-69, Ram:al-94} adapted to take into account a broken power-law electron spectrum.

Figure~\ref{Fig_gyro} (upper panel) shows the results for an electron population with a high energy cutoff $\rm{E}_{\rm{max}}=10,000~\rm{keV}$ and different low-energy cutoffs $\rm{E}_{\rm{min}}=$  30 keV (solid line), 400 keV (dotted line) and 800 keV (dashed line). In Fig.~\ref{Fig_gyro} (lower panel), $\rm{E}_{\rm{min}}=30~Ê\rm{keV}$ and  $\rm{E}_{\rm{max}}=$ 10,000 keV (solid line), 7,000 keV (dotted line) and 4,000 keV (dashed line).   Figure~\ref{Fig_gyro} shows that the optically-thin part of the observed radio spectrum is emitted by electrons in the $\sim$ 800 to 7,000--10,000 keV energy range, that is above the break energy of the electron spectrum deduced from the HXR/GR photon spectrum. This is consistent with earlier findings \citep{Tro:al-98,Tro:al-00,Tro:al-08}. B and T$_{\rm{trap}}$, which fix the value of N$_{\rm{R}}$, are not independent parameters in the data fitting process. Indeed, the same flux densities can be obtained for larger B and lower T$_{\rm{trap}}$ and vice versa. $\rm{B\sim700~G}$ and $\rm{T_{trap}\sim2~s}$ appears to be a reasonable set of values. On one hand, a substantial increase of B would lead to a too high turnover frequency. On the other hand, a decrease of B would imply an increase of $\rm{N_R}$. Since there is no physical reason to have E$_{\rm{min}}$ larger than a few tens of keV, a too low value of B would lead to an unreasonably high density of non-thermal electrons in the radio source compared to $\rm{N_{amb}}$.

 \section{Origin of the 30 THz emission}
      \label{S-30T}

   In the following we assume that both the quiet and flare 30 THz emission is thermal emission from the chromosphere. The aim of this section is to search from which atmospheric layer(s) the 30 THz emission observed during the \event\ flare \citep[see Section~\ref{S-Over} and ][]{Kau:al-13} arises. As we do not know the actual flaring atmosphere, to achieve this, a first approach employs  time-independent models of both the flaring and quiet lower atmospheres of the Sun. Such semi-empirical models are based on hydrostatic and statistical equilibria. For a given temperature structure the full non-LTE radiative transfer is solved numerically  in order to obtain the synthetic flare spectrum. This synthetic spectrum is compared to observed spectral  features (both lines and continuum) and the temperature structure is adjusted in order to get a good agreement with observations. 
\begin{table}
\label{Tab_model}
\caption{Height ( h$_{\rm{max}}$), opacity ($\rm{\tau_{30}}$) at the maximum of CF$_{30}$. The altitudinal range $\Delta\rm{h}$ over which CF is larger than half of its maximum value. $\rm{T_{b30}}$ is the derived brightness temperature at 30 THz. $\Delta$T$_{\rm{b30}}$  is the increase of $\rm{T_{b30}}$ with respect to \CS\ and  $\Delta$S$_{30}$ is the excess  flux density with respect to \CS\ for a 30 THz emitting source of 10\arcsec\ diameter.}
\begin{tabular}{ccccccc}
\textbf{Model}  & h$_{\rm{max}}$     &$\rm{\tau_{30}}$   &  $\Delta\rm{h}$   &  $\rm{T_{b30}}$  & $\Delta$T$_{b30}$  & $\Delta$S$_{30}$\\
                   & [km]      &             &          [km]       &  [K]             &   [K]        & [sfu] \\
\hline
\CS\           & 180       &  1.1     & 130--270       &    4700       &               &   \\
\FLA\         & 180       &  1.1      &  130--270      &    4900      &   200      & 1000\\
\FO\          & 180       &   1.4     &  140--300      &    5000      &   300      & 1500\\
\FT\           & 1070    &  0.2      &  960--1100    &    7100      &   2400    & 12,300\\
 
\end{tabular}
\end{table}
   
In order to compare expectations from flare models with quiet Sun conditions we have used the  \CS\  model  for the quiet solar atmosphere \citep{Avr:Loe-08}. The flare atmospheric models considered hereafter are: the FLA model constructed by \cite{Mau:al-90} to account for the WL continuum emission observed during the SOL1983-06-15 flare, and the F1 and F2  models  representative of respectively weak and medium-size flares  \citep{Mac:al-80}.  We have restricted our analysis to these models because they are well documented in the literature and because \so\ is a medium size flare.Ê We take their temperature height profiles as inputs and use the PAKAL radiative transfer code \citep{Luz:al-10,Luz:al-11},  which is similar to the PANDORA code  used by \cite{Avr:Loe-03}, in order to:  (i) compute the densities of electrons and of the most important ion species, including $\rm{H^-}$, as a function of height h and, (ii)  determine the  absorption coefficient $\rm{\kappa_{30}(h)}$ and the opacity $\rm{\tau_{30}(h)}$ for a 30 THz emitting region at a longitude of 60$^{\circ}$. 
Following \cite{Hei:-12}, we then compute the contribution function $\rm{CF_{30}(h)}$ as a function of h, given by:
\begin{equation}
\mathrm{CF_{30}(h)} = \rm{\eta_{30}(h) e^{-\tau_{30}(h)} \ , \quad 
\eta_{30} (h)= \kappa_{30}(h) B_{30}(T(h)) \ },
\end{equation}
where T(h) is the local temperature and $\rm{B_{30}(T(h))}$ the Planck source function. $\rm{CF_{30}(h)}$ allows us to identify, for each model,  the atmospheric layers which contribute most to the 30 THz emission. For a given model the brightness temperature at 30 THz is given by:
  \begin{equation} 
 \mathrm{T_{b30} = \int T e^{-\tau_{30}} d \tau_{30}} .\ 
   \end{equation}

     \begin{figure}    
 \centerline{\includegraphics[width=\textwidth,clip=]{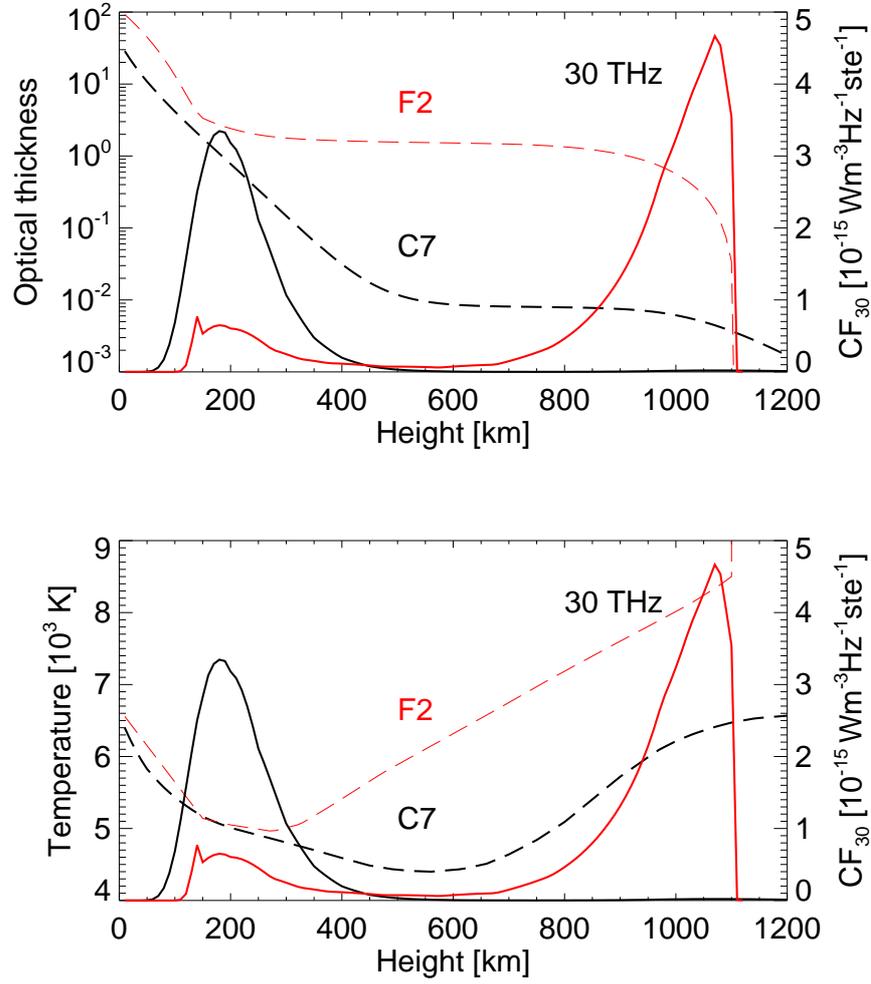}
              }
              \caption{Top: Opacity $\tau_{30}$ (dashed lines) and contribution function CF (solid lines) as a function of height above the photosphere. Red curves represent \FT\ model results, while
black curves are for \CS.  Bottom: Atmospheric local temperature (dashed lines) and
CF (solid lines) as a function of height. }
   \label{Fig_c7f2}
   \end{figure}

For each atmospheric model Table~3 displays  h$_{\rm{max}}$ and $\rm{\tau_{30}}$ at the maximum of $\rm{CF_{30}}$, the altitude range $\Delta\rm{h}$ at half of  the $\rm{CF_{30}}$ maximum,  the expected brightness temperature $\rm{T_{b30}}$, the excess brightness temperature $\Delta\rm{T_{b30}}$ and the corresponding excess flux density $\Delta$S$_{30}$ (for a 10\arcsec\ source) with respect to values obtained for C7. The dashed lines in Figure~\ref{Fig_c7f2} show  $\rm{\tau_{30}}$ (top) and T (bottom)  as a function of h for  \CS\ (black) and \FT\ (red). The corresponding height profiles of CF$_{30}$(h) (solid lines) are over plotted in both panels. 
The 30 THz emission from C7, FLA and F1 arises from an atmospheric layer LC7$_{\rm{l}}$ centered at h$_{\rm{max}}$  $\sim$ 180~ km, where  $\rm{\tau_{30}}\sim 1$,  which extends from $\sim$ 130--140 to 270--300 km . Thus, for  these models, the 30 THz radiation arises from an optically thick layer located below the temperature minimum region at temperatures around 5000 K.  For the quiet Sun, this is consistent with earlier results  \citep[see \textit{e.g.} Figure~2 in ][]{Dem:al-91}.  FLA and F1 cannot explain the 30 THz emission observed  during \so\ because  they lead to $\Delta$S$_{30}$$\sim$ 1000-1500 sfu, which is about one order of magnitude lower than the observed value $\sim$ 12,000 sfu at the maximum of the 30 THz burst. On the contrary, F2 leads to $\Delta$S$_{30}$ $\sim$ 12,300 sfu, which is in agreement with the observed value. In F2, the maximum of  CF$_{30}$ occurs at h$_{\rm{max}}$ $\sim$ 1070 km where $\rm{\tau_{30}}$ $<$  1 so that  the 30 THz radiation arises mostly ($\sim$ 80\%) from an atmospheric layer LF2, $\Delta$h$_{\rm{h}}$ $\sim$ 960-1100 km, located above the temperature minimum region at  temperatures around 8000 K. Figure~\ref{Fig_c7f2}  shows that for \FT, in addition to the LF2 layer, there is an optically thick 30 THz emitting layer, slightly below the temperature minimum, in the altitude range $\Delta h_1 \sim$ 130--270 km. This latter source contributes moderately ($\sim$ 20\%) to the total emission. Its temperature is a few tens of kelvin larger than that of the quiet Sun so that it represents only less than 1\% of the quiet Sun emission. This is in line with earlier theoretical expectations by \cite{Okh:Hud-75} which indicate that the contrast between the quiet and flare infrared emission in the 10-40 THz range is about ten times higher for an optically thin source at about 10$^4$ K than for an optically thick source  around 5000 K. This is also consistent with results of time-dependent numerical simulations \citep[see Figure~3 in][]{Kas:al-09b}. The \FT\ flare model  appears thus as a plausible atmospheric model to account for the 30 THz emission radiated during \so.

 \section{30 THz chromospheric response to flare accelerated particles}
 	 \label{S-Resp}
The 30 THz time profile (see Figure~\ref{Fig_over})  exhibits an impulsive emission, which encompasses the two radio and HXR peaks, superimposed on  a long-lasting and slowly varying component (see Section~\ref{S-Over}). This suggests that the 30 THz impulsive burst results, directly or indirectly,  from chromospheric heating due to energy deposition by non-thermal particles (electrons, protons and $\alpha$ particles) while  the slowly varying emission is generated by some slower process like, \textit{e.g.},  conduction fronts.  In the following we focus on the chromospheric heating by particles accelerated during \so. A proper treatment of this problem, which is beyond the scope of the present study,  should account for the temporal evolution of the flaring atmosphere due to  time dependent heating by particles as in  \cite{Kas:al-09b, Kas:al-09a}. Here, we simply consider C7 as the initial atmosphere model at t$_0$ $\sim$ 17:20:45 UT (beginning of the HXR and 30 THz radiation) and \FT\ as the atmosphere model that is reached at t$_1$ $\sim$ 17:24:50 UT (maximum of the observed 30 THz radiation).  For a given atmosphere model we have calculated the energy deposition per centimeter by non-thermal particles $\Theta(N)$ (erg cm$^{-1}$) as a function of the hydrogen column depth $N$ (see Eq.~\ref{edep3} in the Appendix). In the following we use height $h$ to parametrize altitude above the photosphere in a given atmosphere, considering $\Theta(h)$ rather than $\Theta(N)$. These computations are presented in detail in the Appendix. They  extend previous works  \citep[\textit{e.g.}][]{Ems-78} by considering that the degree of ionization varies with $N$ and retaining the energy-dependence of the Coulomb logarithm, which is particularly important for ions.

\begin{figure}    
 \centerline{\includegraphics[width=\textwidth,clip=]{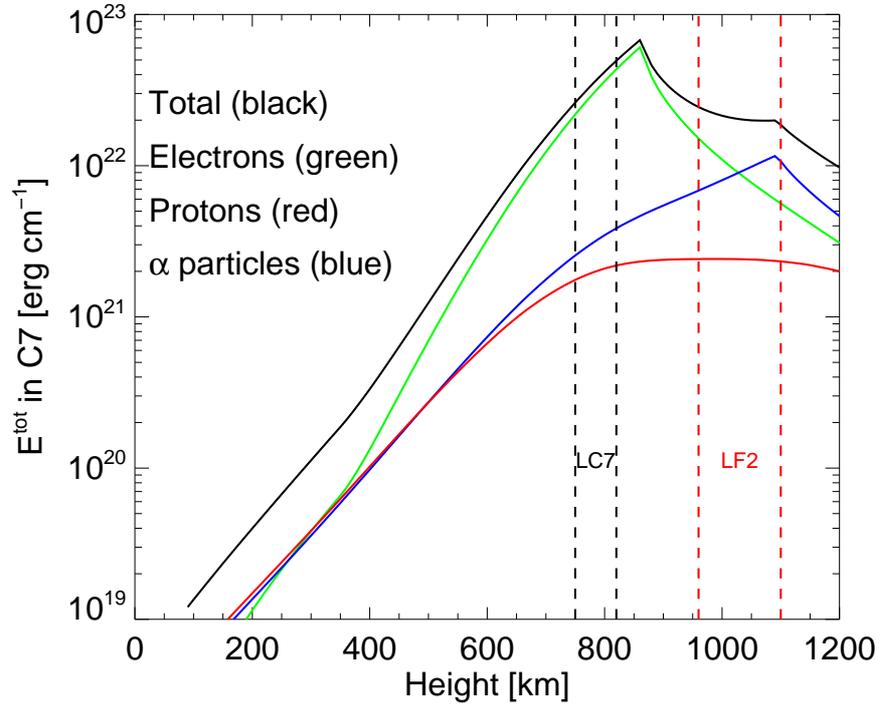}
              }
              \caption{Energy deposition per unit of height in C7 (E$^{\rm{tot}}$) by flare accelerated particles as a function of height. The dashed vertical lines in red show the F2 atmospheric layer (LF2) from which $\sim$~80\% of the 30 THz emission arises. The black vertical lines show the C7 layer (LC7) which covers the same mass column density  as the LF2 layer.}
   \label{fig_edep}
   \end{figure}

As a first approach, we have computed the total energy deposited by particles accelerated between t$_0$ and t$_1$ as a function of $h$  in C7. For that we have integrated the BGO count rate spectrum spectrum from t$_0$ to t$_1$ and estimated the mean flux of electrons, protons and $\alpha$ particles in the same way as in Section~\ref{S-xrgr} and multiplied by t$_1-\rm{t}_0$ = 245 s . This leads to  $\mathcal{F}_{0\rm {e}} (>50~ \rm{keV}) \sim 10^{37}$, $\mathcal{F}_{0\rm{p}}(>\rm{30~MeV}) \sim 8.5 \times 10^{30}$ and $\mathcal{F}_{0\rm{\alpha}}(\rm{>30~MeV/nucl}) \sim 1.9 \times 10^{30}$ for respectively the total numbers of $>$ 50~keV electrons, $>$ 30~MeV protons and $>$ 30~MeV/nucl $\alpha$ particles accelerated between t$_0$ and t$_1$. We then use these normalizations in Eq. (\ref{edep3})  to calculate the energy deposited per cm due to electrons, protons and ions, and sum over these to obtain the total energy deposited by particles per cm   $\rm{E^{tot}}\rm{(h)}$. Figure~\ref{fig_edep} shows $\rm{E}\rm^{tot}{(h)}$ as a function of h in the pre-flare atmosphere C7 for the ensemble of accelerated particles as well as the contributions from electrons, protons and $\alpha$ particles separately.  As stated in Section~\ref{S-30T} most of the 30 THz emission arises from the LF2 layer marked by red vertical dashed lines in Figure~\ref{fig_edep}.  The altitude range of the C7 layer LC7 (750--820 km), which covers the same range  of mass column densities ($3.5 \times 10^{-3}$--$6.2 \times 10^{-3}$  g cm$^{-2}$) as LF2,  is marked by the black dashed lines in the Figure~\ref{fig_edep}. Such a range of column density is consistent with that indicated by \cite{Mac:al-89}. As a rough approximation we assume that the total  energy deposited by particles between t$_0$ and t$_1$ in LC7, E$_{\rm{par}}^{\rm{LC7}}$, may account for  the 30 THz emission from LF2.  By integrating $\rm{E}\rm{^{tot}{(h)}}$ over the LC7 altitude range we get E$_{\rm{par}}^{\rm{LC7}} \sim 3\times10^{29}$ erg.  The total amount of energy radiated at 30 THz between t$_0$ and t$_1$ is given by:
  \begin{equation} 
 \mathrm{E_{30}^{tot} =10^{-15}~4\pi~R^2 \Delta\nu \int_{t_0}^{t_1}\Delta S_{30}(t) dt~\sim~10^{28}~erg} , \ 
   \end{equation}
   where R=1 AU=1.5~$\times$~10$^{11}$ m,  $\Delta\nu$ is the observed frequency bandwidth in Hz (see Section~\ref{S-Inst}) and $\Delta \rm{S_{30}(t)}$ the excess flux density at 30 THz in sfu. E$_{\rm{par}}^{\rm{LC7}}$ is about thirty times larger than E$_{\rm{30}}^{\rm{tot}}$. Thus, direct heating by non-thermal particles appears as a possible mechanism to account for  the $\sim$ 80\% of the impulsive 30 THz emission which arises from LF2. This latter statement remains valid  even if we included a range of initial pitch angles, mirroring and pitch angle scattering in the energy deposition calculation (see~Appendix). On the other hand, the energy ~ 6$ \times 10^{26}$ erg deposited by particles in LC7$_{\rm{l}}$, which corresponds to about the same mass column density and altitude ranges as the equivalent layer in F2 (see Section~\ref{S-30T})   is too small to account for the remaining 20\% of the 30 THz radiation. The heating in the LC7$_{\rm{l}}$ layer may thus be due to some indirect mechanism, such as radiative backwarming of the deep chromosphere from the LF2 layer \citep[\textit{e.g.},][]{Met:al-90,Ker:Fle-14} or radiative coupling of the upper chromosphere and temperature minimim regions \citep{Mac:al-89}. It should be noted that  a harder ion distribution (\textit{e.g.} $\delta = 2$) would deposit a greater fraction of its total energy content in the deeper atmosphere. To this extent the calculations here, with a fixed value of $\delta = 4$ should be regarded as illustrative. The statistical quality of the $\gamma$-ray spectrum for this event does not justify more detailed investigation, however.
   
Figure~\ref{Fig_over} shows that the evolution of the H$\alpha$ emission from the S$_1$ and S$_2$ kernels (see Figure~\ref{Fig_images}) is similar to that of the 30 THz excess flux density. Because the H$\alpha$ emission arises from a $\sim$ 10$^4$ K plasma, the altitudes of S$_1$ and S$_2$ should be close to that of the 30 THz source at $\sim$ 8000 K. This suggests that direct heating by particles is also responsible for the impulsive  H$\alpha$ radiation while, like for the 30 THz emission, the slowly-varying, long-lasting component is due to some other process. A detailed and quantitative analysis of the WL emission is beyond the scope of this study. Nevertheless, as remarked in Section~\ref{S-Over}, the WL time profile exhibits only the impulsive emission and not the slowly-varying, long-lasting component. This suggests that the mechanism responsible for the 30 THz and H$\alpha$ slowly-varying, long-lasting component is not efficient enough to produce WL emission whether it arises 
 from direct particle heating from a LF2-like layer or from indirect heating of the deep chromosphere (LC7$_{\rm{l}}$  layer).  The former hypothesis is consistent with the heights of WL sources, $\sim$ 800--1000 km above the photosphere estimated by \textit{e.g.}  \cite{Kru:al-15} for some flares. 


 \section{Summary}
       \label{S-Sum}
The \so\ flare is the first event that have been detected in the mid-infrared domain at 30 THz \citep{Kau:al-13}.  The combined analysis of the 30 THz data with HXR/GR, WL, H$\alpha$ and 1-400 GHz radio observations has allowed us to estimate the numbers of electrons, protons and $\alpha$ particles accelerated during the flare and  to discuss the origin of the 30 THz emission. The main findings can be summarized as follows:
\begin{itemize}
\item [-] The HXR/GR observations of the Gamma-ray Burst Monitor  on {\it FERMI} have revealed that \so\ is a gamma-ray line flare detected up to at least 10 MeV.  The numbers and energy contents of HXR/GR emitting electrons, protons and $\alpha$  particles have been estimated. The energy content in protons and $\alpha$ particles with energies above 1 MeV/nucl. is found to be similar to the energy content in $>$ 50 keV electrons. This indicates that, in addition to electrons, protons and $\alpha$ particles may substantially contribute to the heating of the chromosphere during flares such as \so.
\item [-] The observed 1--200 GHz radio spectrum is reminiscent of  gyrosynchrotron radiation. The  optically-thin part of the radio spectrum ($>$ 10--20 GHz) is radiated by HXR/GR emitting electrons with energies greater than $\sim$ 800 keV. The spectrum of these high energy electrons is harder than that of lower energy electrons.
\item [-] Assuming that the 30 THz emission is thermal emission from the chromosphere we have used time independent semi-empirical models of both the quiet and flaring atmospheres in order to identify  from which atmospheric layers the 30 THz emission arises. In agreement with earlier works we find that the quiet Sun 30 THz radiation is emitted by an optically thick layer, at temperatures around 5000 K, below the temperature minimum region. During the flare, the measured flux density excess at 30 THz is consistent with that expected from the F2 model by \cite{Mac:al-80} which is representative of medium size flares. In this model, about 80\% of the 30 THz emission arises from an atmospheric layer LF2 ($\sim$ 960--1100 km above the photosphere) at temperatures around 8000 K well above the temperature minimum region. At 30 THz, this layer is not optically thick ($\tau < 1$). The remaining 20\% of the 30 THz emission arises from an optically-thick layer, slightly below the temperature minimum region in F2, which covers similar altitude and mass column density ranges  to those of the emitting layer in the quiet Sun.
\item [-] The energy deposited by electrons, protons and $\alpha$ particles in the quiet Sun layer which covers the same range of mass column density (3.5--6.2 $\times 10^{-3}$ g~cm$^{-2}$) has been calculated by taking into account that the degree of ionization varies with height and by retaining the energy-dependence of the Coulomb logarithm. This energy is found to be about thirty times greater than that radiated at 30 THz. Thus energy deposition by accelerated particles is a plausible mechanism responsible for the heating of the LF2 layer which gives rise to $\sim$~80\% of the observed 30 THz emission during the impulsive phase of \so. On the other hand, the energy deposited by particles in the deep chromosphere is too small to account for the remaining 20\% of the 30 THz flux density radiated from below the temperature minimum region in F2. A
  possible, but not unique, scenario is that the impulsive WL emission
  and 20\% of the 30 THz radiation arises from an optically-thick
  layer, below the temperature minimum region, heated by
  \textit{e.g.} backwarming from the upper chromosphere where most of the
  energy transported by accelerated particles is deposited.

\item [-] The 30 THz and H$\alpha$ exhibit a long-lasting and slowly
  varying component which is not observed in the WL and HXR/GR
  domains. This indicates that the chromospheric heating giving rise
  to this component is due to a different mechanism than energy
  deposition by accelerated particles and that the efficiency of this
  mechanism is not sufficient to produce long-lasting WL emission. 
 \end{itemize}
 
In summary, using time-independent semi-empirical models of the quiet
and flare atmosphere, we have shown that thermal emission from the
chromosphere heated by accelerated particles provides a plausible
quantitative interpretation of the 30 THz emission in gross agreement
with that observed during the impulsive phase of \so.  Time dependent
numerical models of energy deposition by accelerated particle such as
developed by \cite{Kas:al-09b} are needed to confirm our
conclusion. Moreover, if the 30 THz emission arises from a quasi
optically-thin chromospheric layer like LF2, the emission spectrum
between $\sim$ 10--100 THz is expected to be rather flat or even
slightly decreasing with increasing frequencies \citep[see Figure~2
  in][]{Okh:Hud-75}.  Observations at frequencies above 30 THz such as
those recently obtained from the \textit{McMath-Pierce Solar Facility} at Kitt
Peak \citep{Pen:al-15} should allow us to test the validity of such an
expectation, and thus of the thermal interpretation of the infrared
emission from flares proposed in this work. 
In the near future, the combination of  McMath observations and   
 measurements at   3 and 7 GHz by \textit{Solar-T}  \citep{Kau:al-14} should allow to 
 complete our understanding of the chromospheric response to flare energy release.

\begin{acks}
 The authors thank G. Chambe and K.-L. Klein for their suggestions and critical comments.  We thank STFC for support through grant ST/L000741/1 (ALM).  Some of ALM contribution was carried out while on study leave at CRAAM, Mackenzie Presbyterian University, S\~ao Paulo with FAPESP financial support. PJAS acknowledges the European Community's Seventh Framework Programme (FP7/2007-2013) under grant agreement no. 606862 (F-CHROMA) for financial support. VDL thanks Catedras-CONACyT project 1045. This research was partially supported by Brazil agencies FAPESP (contract 2013/24155-3),CNPq (contract 312788/2013-4), Mackpesquisa and U.S. AFOSR. We are grateful to the referee, S\"am Krucker, for his constructive recommendations. 
\end{acks}

\appendix
\label{app}
In order to compute the energy deposited by accelerated electrons, protons and $\alpha$ particles  in the chromosphere as a function of depth, we inject particles with an energy spectrum  $F_0(E)$ (MeV$^{-1}$) at the top of the atmosphere. We parametrize position in the atmosphere using the \emph{hydrogen} column depth $N$. The energy deposited (erg cm$^{-1}$) at depth $N$ is \citep[following ][]{Ems-78}
\begin{equation}
\Theta(N) \, = \, \int_{E_{min}(N)}^\infty \frac{F_0(E_0)}{v(E)} \left| \frac{\mathrm{d}E}{\mathrm{d}t}(E,N)\right| \mathrm{d}E_0 .
\label{edep1}
\end{equation} 
Here $E_{min}(N)$ is the minimum energy of particle that can reach depth $N$. In Equation (\ref{edep1}) $E$ should be understood as depending on $E_0$ and $N$: it is the energy that a particle of initial energy $E_0$ has when it has reached depth $N$. 

An ion of energy $E$ and charge $z$ loses energy with depth at a rate given by
\citep{Gou-72a, Ems-78, Oli:al-14}
\begin{equation}
\frac{\mathrm{d}E}{\mathrm{d}N} \, = \, -\frac{K}{\beta^2} \Lambda_{eff}(E,N)
\label{eloss}\end{equation}
\noindent Here $\beta$ is particle speed in units of the speed of light, $K$ is given by
\begin{displaymath}
K = \frac{4 \pi e^4}{m_ec^2} z^2  \, ,
\end{displaymath}
\begin{equation}
\Lambda_{eff}(E,N) \, = \, x \Lambda + (1-x) \Lambda' ,
\end{equation}
\noindent
where $x$ is the degree of ionisation and $\Lambda$ and $\Lambda'$ are the Coulomb logarithms appropriate to slowing on electrons, and on neutral atoms respectively. For ions, expressions for $\Lambda$ are given in \textit{e.g.} \cite{But:Buc-62, Gou-72b}. For $\Lambda'$ we used the Bethe-Bloch form \citep{Oli:al-14}, neglecting the correction for the polarisation of the medium which is negligible in the low-density conditions of the solar atmosphere. We estimate the effects of the atmospheric chemical composition on energy loss by multiplying $\Lambda'$ by 
\begin{displaymath}
\sum_n \frac{a_n}{a_H} \frac{Z_n}{A_n} \, = \, 1.16
\end{displaymath}
\noindent where $n$ sums over atomic species and $A_n$, $Z_n$ and $a_n$ are the atomic number, atomic weight and abundance relative to hydrogen of species $n$. Also we generalise $x = n_e/n_H$ so that it is no longer quite the ``degree of ionisation'' and may take values $>1$, dependent on the local electron density.

$\Lambda$ and $\Lambda'$ for electrons are given in \cite{Gou-72b, Ems-78}. In using only Equation (\ref{eloss}) we neglect pitch-angle scattering. Scattering in angle is unimportant for fast ions and relativistic electrons - the only particles that can reach the LC7$_{\rm{l}}$ layer. For non-relativistic electrons it reduces the mean range by 2/3 \citep{Bro-72}, although dispersion means that electrons will stop and deposit energy over a range of depths up to their full vertical stopping depth \citep{Bai-82}. For the present purposes of estimation we neglect pitch-angle scattering completely.

\begin{figure}
\centering
\centerline{\includegraphics[width=0.75\textwidth,clip=]{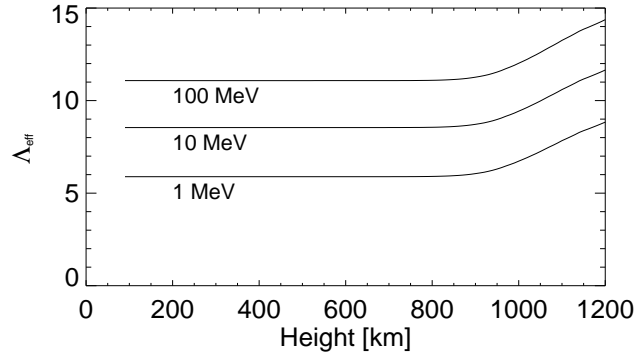}}
\caption{The effective Coulomb logarithm, $\Lambda_{eff}$ through the atmosphere for protons of energy 1, 10 and 100 MeV.}
\label{lamplot}
\end{figure}

Figure~\ref{lamplot} shows the value of $\Lambda_{eff}$ throughout the chromospheric portion of the C7 atmosphere of \cite{Avr:Loe-03}, for protons of energies 1, 10 and 100 MeV. The importance of retaining the energy-dependence of $\Lambda$ has been emphasised elsewhere \citep{Mac:Ton-03}. The variable degree of ionisation means that $\Lambda_{eff}$ also depends strongly on position. Thus the energy loss rate of Equation (\ref{eloss}) is not separable in energy and depth, and analytical heating rates like those of \cite{Ems-78} may not be used (although an energy loss rate appropriate to a completely neutral atmosphere would give a good approximation below about 1000 km). 

To evaluate the energy deposition $\Theta$ we need to assume a form for $F_0(E_0)$, \textit{e.g.} a single power law:
\begin{equation}
F_0(E_0) \, = \, \frac{\mathcal{F}_0}{E_*} (\delta-1) \left(\frac{E_0}{E_*}\right)^{-\delta} , 
\label{eff0}
\end{equation}
\noindent where $\mathcal{F}_0$ is the number of particles injected above energy $E_*$. Then applying Equation (\ref{eloss}) in Equation (\ref{edep1}) gives
\begin{equation}
\Theta(N) \, = \, 8.16 \times10^{-31} z^2 n_H(N) \mathcal{F}_0 E_*^{\delta-1}(\delta-1) \int_{E_{min}(N)}^\infty E^{-\delta} \frac{\Lambda_{eff}(E,N)}{\beta^2(E)}  \mathrm{d}E_0
\label{edep3}
\end{equation} 
Again, $E$ in the integrand of Equation (\ref{edep3}) is understood to be a function of $E_0$ and $N$: the energy at depth $N$ of a particle that had energy $E_0$ at injection. Since we cannot obtain analytical expressions for $E(E_0,N)$ we calculate it via numerical integration of Equation (\ref{eloss}), with numerical interpolation as necessary between the tabulated points of a given  atmosphere model, as needed for each of the abscissae of a numerical evaluation of the energy deposition (Equation (\ref{edep3})). More elaborate forms of $F_0(E_0)$, \textit{e.g.} the broken power-law deduced for electrons, are straightforwardly substituted in Equation (\ref{edep1}) as necessary.

\bibliographystyle{spr-mp-sola}
\tracingmacros=2
\bibliography{bib_trottet_et_al_20120313-new}  

 \IfFileExists{\jobname.bbl}{} {\typeout{}
\typeout{****************************************************}
\typeout{****************************************************}
\typeout{** Please run "bibtex \jobname" to obtain} \typeout{**
the bibliography and then re-run LaTeX} \typeout{** twice to fix
the references !}
\typeout{****************************************************}
\typeout{****************************************************}
\typeout{}}

\end{article} 

\end{document}